\documentclass[preprint,amsmath,titlepage,amssymb,english,aps,showkeys]{revtex4}
\usepackage{xcolor}
\usepackage{amssymb,amsmath}
\usepackage{graphicx}
\usepackage{hyperref}
\usepackage[utf8]{inputenc}
\usepackage[T1]{fontenc}
\usepackage{amsmath}
\usepackage{amsfonts}

\begin{document}

\title{Perturbative analysis of singularity-free cosmological solutions \\ in unimodular Kaluza-Klein theory}

\author{J{\'u}lio C. Fabris}
\email{julio.fabris@cosmo-ufes.org}
\affiliation{Núcleo Cosmo-ufes\&Departamento de F{\'i}sica, CCE, Universidade Federal do Espírito Santo, Vitória, ES, Brazil}

\author{St\'efani Faller}
\email{stefani.ganda@edu.ufes.br}
\affiliation{Núcleo Cosmo-ufes\&Departamento de F{\'i}sica, CCE, Universidade Federal do Espírito Santo, Vitória, ES, Brazil}

\author{ Richard Kerner}
\email{richard.kerner@sorbonne-universite.fr}
\affiliation{Laboratoire de Physique Th\'eorique de la Mati\`ere Condens\'ee, Sorbonne-Universit\'e, Boîte 121, 4 Place Jussieu, 75005, Paris,  France}

\begin{abstract}

The unimodular version of the Kaluza-Klein theory is briefly recalled, and its projection on the $4$-dimensional spacetime is constructed.
Imposing unimodularity condition on the $5$-dimensional Kaluza-Klein metric, det$g_{AB}=1$ is equivalent with introducing cosmological
term in Einstein's equations in $4$ dimensions, and with scalar field of the Brans-Dicke type. 
Singularity-free cosmological solutions with scalar field and with matter sources are constructed, and their basic properties analyzed,
along the results obtained in our previous publications (see \cite{FabrisKerner2024}, \cite{FabrisKerner2025}).
In the present paper, attention is focussed on the perturbative analysis of cosmological solutions, providing a clue concerning their
stability against small fluctuations. 

\end{abstract}


\maketitle

\title{Perturbative analysis of singularity-free cosmological solutions \\ in unimodular Kaluza-Klein theory}

\author{J\'ulio C. Fabris$^a$, St\'efani Faller$^a$ and Richard Kerner$^b$}

\address{$^a$ F\'isica Teorica, Universidade Federal do Espirito Santo, \\ Vit\'oria, Brazil \\
$^b$: Laboratoire de Physique Th\'eorique de la Mati\`ere Condens\'ee \\
Sorbonne-Universit\'e, 4 Place Jussieu, 75005 Paris, France}


\section{Introduction}
\vskip 0.3cm
\indent
The Unimodular Gravity was considered already in $1919$ by Einstein (\cite{ein}), who investigated an alternative to the initial 
formulation of General Relativity by imposing the extra unimodularity condition on the pseudo-Riemannian metric, $\mid {\rm det} g_{\mu \nu} \mid = 1$.
This condition seemed natural for at least two reasons: firstly, the Minkowskian metric tensor $\eta_{\mu \nu} = {\rm diag} (+1,-1,-1,-1)$
has this property, and its infinitesimal deformations $h_{\mu \nu}$ should be tracelss if the unimodularity was to be also imposed on 
the deformed metric $g_{\mu \nu} + \epsilon h_{\mu \nu}$; secondly, it permitted to replace the cosmological term by the Lagrange multiplier 
taking into account this constraint in the variational principle. In $1919$ the Friedmann solution was not known yet, and Einstein \cite{Einstein1917} was 
attached to the Aristotelean vision of eternal and stationary Universe (at a very large scale, of course). This is why he introduced the cosmological 
term $\Lambda g_{\mu \nu}$ on the right-hand side of the field equations, acting as a source of negative pressure ensuring the balance 
with non-zero matter density and its repulsive contribution to gravity. 

Closer to present times, the Unimodular Gravity is considered as an alternative, and very promising, approach to the cosmological constant problem, 
see Refs. \cite{Henneaux,Dragon, Unruh,wein}.

Let us show how the unimodularity of the metric modifies the Einstein-Hilbert action integral using this extra constraint
as a Lagrange multiplier technique in the corresponding variational principle, and how the new Einstein equations look like. 
For alternative approaches, see Refs. \cite{ea,u1}.

Let us consider the action
\begin{eqnarray}
{\cal S} = \int d^4x\biggr\{\sqrt{-g}R - \lambda(\sqrt{-g} - \xi)\biggl\} + \int d^4 x \sqrt{-g}{\cal L}_m.
\end{eqnarray}
In this expression, $\lambda$ is a Lagrange multiplier, and $\xi$ is an external field introduced in order to give flexibility in the choice of the coordinate system. 
As a special case, $\xi$ may be considered a pure number, which implies the choice of specific coordinates.  

The variation with respect to the metric leads to:
\begin{eqnarray}
\label{erg1}
R_{\mu\nu} - \frac{1}{2}g_{\mu\nu}R + \frac{\lambda}{2}g_{\mu\nu} = 8\pi GT_{\mu\nu}.
\end{eqnarray}
The variation with respect to $\lambda$ leads to the unimodular constraint: 
\begin{eqnarray}
\label{vin-rg-1}
\xi = \sqrt{-g}.
\end{eqnarray} 

The trace of (\ref{erg1}) implies:
\begin{eqnarray}
\lambda = \frac{R}{2} + 8\pi G\frac{T}{2}.
\end{eqnarray}
Inserting this result in (\ref{erg1}), we obtain the unimodular field equations:
 \begin{eqnarray}
 \label{erg2}
 R_{\mu\nu} - \frac{1}{4}g_{\mu\nu}R = 8\pi G\biggr(T_{\mu\nu} - \frac{1}{4}g_{\mu\nu}T\biggl).
 \end{eqnarray}
The theory is now invariant by a restricted class of diffeomorphisms, often called ``transverse diffeomorphisms''. 
 
The Bianchi identities imply that the usual energy-momentum tensor conservation must be generalized as follows:
 \begin{eqnarray}
 \label{brg1}
 \frac{R^{;\nu}}{4} = 8\pi G\biggr({T^{\mu\nu}}_{;\mu} - \frac{1}{4}T^{;\nu}\biggl).
 \end{eqnarray} 

If the usual energy-momentum tensor conservation is imposed (which in the unimodular context, 
would be in opposition with basic General Relativity theory), 
 \begin{eqnarray}
 \label{cons-rg-1}
 {T^{\mu\nu}}_{;\mu} = 0,
 \end{eqnarray}
 the equation (\ref{brg1}) becomes
 \begin{eqnarray}
 \label{brg2}
 \frac{R^{;\nu}}{4} = -2 \pi GT^{;\nu}.
 \end{eqnarray} 
The above equation (\ref{brg2}) may be integrated leading to, 
\begin{eqnarray}
\label{ic1}
R = - 2\pi GT + 4\Lambda,
\end{eqnarray}
where $\Lambda$ is an integration constant which may be interpreted as the cosmological constant. 
Inserting the relation (\ref{ic1}) in (\ref{erg1}), the resulting equations are: 
\begin{eqnarray}
 \label{erg3}
 R_{\mu\nu} - \frac{1}{2}g_{\mu\nu}R = 8\pi GT_{\mu\nu} + g_{\mu\nu}\Lambda.
 \end{eqnarray}
We see that the equations of standard General Relativity are recovered, but with a cosmological constant which was absent in the original structure, but
appears naturally as a constant of integration. 

It is worthwhile to remind here that even when the unimodular constraint is satisfied, det $g = 1$, the metric tensor is not constant. Moreover, its
its transformation properties are typical of a {\it tensor density}:
\begin{equation}
g({\tilde{x}}) = {\rm det} \left( \frac{\partial x^{\lambda}}{\partial {\tilde{x}}^{\mu}} 
\frac{\partial x^{\rho}}{\partial{\tilde{x}}^{\nu}}  \right) g(x).
\label{gtransform}
\end{equation}
Thus, 
\begin{equation}
\sqrt{g} ({\tilde{x}}) = ({\rm det}^{-1} J ) \; \sqrt{g} (x),
\label{gtransJ}
\end{equation}
where $J =$ det $( \frac{\partial {\tilde{x}}}{\partial x} )$
A scalar density os a scalar field multiplied by $\sqrt{g}$:
\begin{equation}
{\cal{F}} (x) = F (x) \; \sqrt{g}.
\label{densityF}
\end{equation}
Its covariant derivative with respect to connection $\Gamma^{\lambda}_{\mu \nu}$ is then
\begin{equation}
\nabla_{\lambda} {\cal{F}} = \sqrt{g} \nabla_{\lambda} {F} + F \nabla_{\lambda} {\sqrt{g}} = \partial_{\lambda} F - F \Gamma^{\nu}_{\nu \lambda}.
\label{covdiffF}
\end{equation} 
and the Lie derivative with respect to an arbitrary vector field $X^{\mu}$ is:
\begin{equation}
{\cal{L}}_X {\cal{F}} = \sqrt{g} {\cal{L}}_X {F} + F {\cal{L}}_X {\sqrt{g}} = X^{\mu} \partial_{\mu} {\cal{F}} + {\cal{F}} \partial_{\mu} X^{\mu} 
= \nabla_{\mu} {\cal{F}} X^{\mu},
\label{LieF}
\end{equation}
because
\begin{equation}
\nabla_{\mu} \sqrt{g} = \partial_{\mu}\sqrt{g} - \sqrt{g} \Gamma^{\lambda}_{\lambda \mu} = 0.
\label{covdiffg}
\end{equation}
with respect to the Christoffelian connection, and because
\begin{equation}
\frac{\partial \sqrt{g}}{\partial g_{\mu \nu}} = \frac{1}{2} \sqrt{g} g^{\mu \nu}.
\label{derg}
\end{equation}
we get
\begin{equation}
{\cal{L}}_X {\sqrt{g}} = \frac{\sqrt{g}}{2} g^{\mu \nu} \left( ({\cal{L}}_X {g})_{\mu \nu} \right) = \sqrt{g} \partial_{\mu} X^{\mu}.
\label{Liesqrtg}
\end{equation}
These properties of covariant and Lie derivatives of densities are crucial in the derivation of modified Einstein equations from variational
principle, whose integrand is by definition a scalar density.

\section{Unimodular Kaluza-Klein theory}

The first attempt to unify the new theory of gravitation, proposed by Einstein in $2015$, with Maxwell's theory of electromagnetism, was made by Th. Kaluza
\cite{Kaluza1921} in $1921$. The idea was to extend the General Relativity to $5$ dimensions, where the metric tensor $g_{AB}, \; {\small A,B,...} = 1,2,...5$
can accomodate $15$ field degrees of freedom instead of the $10$ in the $4$-dimensional spacetime. 

Quite obviously, the $15$ independent components of a $5$-dimensional manifold's metric tensor can accomodate not only the $10$ components
of a $4$-dimensional subspace's metric, identified with the metric tensor of General Relativity, but also the extra
four components of the $4$-potential of Maxwell's electromagnetism, if they are identified with mixed components of the $5$D metric:
${\tilde{g}}_{5 \mu} = {\tilde{g}}_{\mu 5}$, where $\mu, \nu = 0,1,2,3$:
\begin{equation}
{\tilde{g}}_{AB} = \begin{pmatrix} g_{\mu \nu} & A_{\mu} \cr A_{\nu} & 1 \end{pmatrix}.
\label{gKK}
\end{equation} 
( In what follows, we shall denote by tilded symbols geometrical objects
defined in the $5$-dimensional Kaluza-Klein space, the same symbols relative to the $4$-dimensional space-time being expressed as usual,
by non-tilded letters.)

In the original version of his model, Kaluza has fixed the $g_{55}$ component equal to $1$, leaving just the $14$ degrees of freedom, but at the
same time taking the risk of getting an over-determined system. Indeed, the standard variational principle applied to the Einstein-Hilbert 
lagrangian in $5$ dimensions leads to $15$ partial differential equations although we have only $14$ independent fields identified
as $g_{\mu \nu}$ and $A_{\mu}$. 

However, by a happy coincidence, even with this incomplete version, the system was not over-determined due to the fact
that out of the $15$ Einstein equations in vacuo corresponding to the components of symmetric Ricci and metric tensors:
$(\mu \nu), \; \; (\mu 5 )$ and $(5 5)$ the last one ${\tilde{R}}_{55} - \frac{1}{2} {\tilde{g}}_{55} {\tilde{R}}$ reduces to tautology $0 = 0$, 
leaving exactly $14$ equations, which are easily recognized as the usual $4$-dimensional Einstein's equations with electromagnetic energy-momentum
tensor as a source, along with Maxwell's equations coupled with gravitational field through covariant derivatives: the $15$ equations
\begin{equation}
{\tilde{R}}_{AB} - \frac{1}{2} {\tilde{g}}_{AB} {\tilde{R}} = 0, \; \; (A, B,..) = (\mu, 5), \; \; 
{\tilde{R}} = R - \frac{1}{4} F_{\lambda \rho} F^{\lambda \rho}.
\label{fiveeqs}
\end{equation}
with an extra assumption that the fields $g_{\mu \nu}$ and $A_{\mu}$ depend exclusively on space-time variables $x^{\mu}$ and not on $x^5$, 
when explicited give rise to the following system of coupled equations:
\begin{equation}
R_{\mu \nu} - \frac{1}{2}g_{\mu \nu} R = g^{\lambda \rho} F_{\mu \lambda} F_{\nu \rho} - \frac{1}{4} g_{\mu \nu} F_{\lambda \rho} F^{\lambda \rho},
\label{KKmunu}
\end{equation}
\begin{equation}
g^{\mu \nu} \nabla_{\mu} F_{\nu \lambda} = 0, 
\label{Maxwellcov}
\end{equation}
the last $55$ component reducing to $0 = 0$. This circumstance is often called ``the Kaluza-Klein miracle''.

But it is not the unique ``bonus'' here. There is another happy coincidence, namely, the fact that the determinant of the $5$-dimensional Kaluza-Klein
metric does not depend on the electromagnetic fields $A_{\mu}$, but is a product of the determinant of the $4$-dimensional metric of the
space-time multiplied by a function of the scalar field. This circumstance encourages the idea of introducing the unimodular condition in the
$5$-dimensional Kaluza-Klein space. 

In its first version proposed by Th. Kaluza, the fifth dimension was just an extra space coordinate, the
entire space being isomorphic with $M_4 \times R^1 \sim [ct, x, y, z, x^5]  \sim M_5$, a five-dimensional Minkowski space. Later on, in $1926$, O. Klein (\cite{Klein1926})
introduced a modification inspired by the freshly discovered quantum theory, which postulated wave-particle duality, according to which elementary particle's
energy $E$ and momentum ${\bf p}$ can be also identified with frequency $\omega$ and wave vector ${\bf k}$ of the corresponding wave.
In the Schroedinger representation to a state of quantum particle corresponds its (complex) wave function $\psi (t, {\bf x})$, and the energy and momentum are 
represented by hermitian operators $ E \rightarrow -i \hbar \partial_t$, ${\bf p} \rightarrow -i \hbar {\boldsymbol{\nabla}}$.

Klein proposed to consider a compact fifth dimension, a circle with a very small radius. 
Supposing that wave functions describing electrically charged point particles are defined on the Kaluza-Klein space, they should depend also on the fifth coordinate $x^5$,
the dependence being periodic if the $5$-th dimension is circular, i.e. isomorphic with a unit circle $S^1$. The most general form of such wave functions should be then:
\begin{equation}
f(x^{\mu}, x^5) = {\displaystyle{\sum_{k=0}^{\infty} }} a_k (x^{\mu}) e^{ik q x^5}.
\label{eik}
\end{equation}
 with dim(q) =cm$^{-1}$.

Identifying the fifth component of momentum $p_5$ with the operator of electric charge, $p_5 \rightarrow - i \hbar \partial_5$ we shall get the
eigenvalues of $p_5$ to be multiples of $q$, identified with the discrete values of electric charge of elementary particles known at that time, electrons and protons.

The unimodularity condition in the Kaluza-Klein case becomes particularly natural in a non-holonomous local frame system. We can define the $5$D spacetime
metric tensor following the tetrad technique known from General relativity: the Riemannian metric is defined as $g_{\mu \nu} = \eta_{ab}\theta^a_{\mu} \theta^b_{\nu}$,
with $\eta_{ab} =  {\rm diag} (+,-,-,-)$, $a, b = 1,..4$, $\mu, \nu = 0,1,2,3$. The dependence of $g_{\mu \nu}$ on space-time coordinates is encoded in local $1$-forms
$\theta^a_{\mu} (x^{\lambda})$.

The non-holonomic local frame is defined by the following choice of local basis of $1$-forms:
\begin{equation}
\theta^{\mu} = d x^{\mu}, \; \; \; \; \theta^5 = dx^5 + q \; A_{\mu} dx^{\mu},
\label{thetas}
\end{equation}
the constant $e$ must have the dimension of length in order to ensure the uniformity with space variables $x^{\mu}$.  

The dual vector fields satisfy $\theta^A ({\cal{D}}_B) = \delta^A_B$:
\begin{equation}
{\cal{D}}_{\mu} = \partial_{\mu} - q \; A_{\mu} \partial_5, \; \; \; \; {\cal{D}}_5 = \partial_5.
\label{CalD}
\end{equation}
Introducing transition matrices $U^A_B$ and ${\overset{-1}{U^B_C}}$ such that $\theta^A = U^A_B dx^B, \; {\cal{D}}_C = {\overset{-1}{U^D_C}} \partial_D $
we can write:
$$U^{\mu}_{\nu} = \delta^{\mu}_{\nu}, \; \; \; U^{\mu}_5 = 0, \; \; U^5_{\mu} = q A_{\mu}, \; \; U^5_5 = 1;$$
\begin{equation}
{\overset{-1}{U^{\mu}_{\nu}}} = \delta^{\mu}_{\nu}, \; \; {\overset{-1}{U^{5}_{\nu}}} = - q A_{\nu}, \; \; {\overset{-1}{U^{\mu}_{5}}} = 0 
\; \;  {\overset{-1}{U^{5}_{5}}} = 1.
\label{UUinv}  
\end{equation}
Quite obviously, both determinants are equal to $1$: det(($U^A_B) = 1$ and det (${\overset{-1}{U^B_C}}) = 1.$ 
The metric tensor expressed in the non-holonomic frame is easily deduced from the square of the  $5$-dimensional length element,
taking on the following form:
\begin{equation}
ds^2 = g_{\mu \nu} dx^{\mu} dx^{\nu} -  \left[ dx^5 + q \; A_{\mu} dx^{\mu} \right] \left[ dx^5 + q \; A_{\nu} dx^{\nu} \right]
\label{deessquare}
\end{equation}
leading to the following $5 \times 5$ matrix representation:
\begin{equation}
{\tilde{g}}^{AB} = \begin{pmatrix} g_{\mu \nu} + q^2 A_{\mu} A_{\nu} & - q A_{\nu} \cr - q  A_{\mu} & - 1 \end{pmatrix} 
\label{gABfive}
\end{equation}
The inverse matrix becomes then:
\begin{equation}
{\tilde{g}}^{BC} = \begin{pmatrix} g^{\nu \lambda}  &  q A_{\lambda} \cr q  A_{\nu} & q^2 A^{\nu} A^{\lambda} - 1 \end{pmatrix} 
\label{invgABfive}
\end{equation}
One easily checks that 
$${\tilde{g}}_{AB} {\tilde{g}}^{BC} = \delta^A_C.$$
In order to make the theory complete, the full set of $15$ field degrees of freedom should be included. The missing scalar field is displayed as the
$55$ component of the Kaluza-Klein metric tensor. The improved version of Kaluza-Klein theory, including scalar field along with gravity and electromagnetism,
was proposed by P. Jordan \cite{Jordan} and Y. Thiry \cite{Thiry}, who discussed also the impact of scalar field on gravity and electromagnetism.

In the complete theory the diagonal metric is chosen to be $g_{AB} =$ diag $(1, -1, -1, -1, - \phi^2(x^{\mu})$, and the local frames and the metric are modified 
consequently, leading to the following form of the metric:
\begin{equation}
ds^2 = g_{\mu \nu} dx^{\mu} dx^{\nu} - \Phi^2 \left[ dx^5 + q \; A_{\mu} dx^{\mu} \right] \left[ dx^5 + q \; A_{\nu} dx^{\nu} \right]
\label{deessquare2}
\end{equation}
leading to the following $5 \times 5$ matrix representation:
\begin{equation}
g^{AB} = \begin{pmatrix} g_{\mu \nu} + q^2 \Phi^2 A_{\mu} A_{\nu} & - q \Phi^2 A_{\nu} \cr - q \Phi^2 A_{\mu} & - \Phi^2 \end{pmatrix} 
\label{gABfive2}
\end{equation}
The inverse matrix becomes then:
\begin{equation}
g^{BC} = \begin{pmatrix} g^{\nu \lambda}  &  q A_{\lambda} \cr q  A_{\nu} &  q^2 A^{\nu} A^{\lambda} - \Phi^2 \end{pmatrix} 
\label{invgABfive2}
\end{equation}

The extension of the Kaluza-Klein structure to include non-abelian gauge fields is described in Ref. \cite{Kerner1981}.

\section{General expressions}

It is possible to construct an effective theory in four dimensions resulting from the Kaluza-Klein Unimodular Gravity (KK-UG). The derivation is sketched in what follows.
The main goal of the present work is to investigate the features of singularity-free cosmological solutions in the effective KK-UG in four dimensions, 
in presence of the matter components, which emerges from the dimensional reduction of the 5 dimensional theory. This framework was settled out in 
Ref. \cite{FabrisKerner2024}, and some simple cosmological configurations were studied in Ref. \cite{FabrisKerner2025}.

The unimodular equations in $D = 5$ dimensions read,
\begin{eqnarray}
\tilde R_{AB} - \frac{1}{5}\tilde g_{AB}\tilde R &=& 8\pi G\biggr\{\tilde T_{AB} - \frac{1}{5}\tilde g_{AB}\tilde T\biggl\},\\
\frac{3}{10}\tilde R_{;A} &=& 8\pi G\biggr\{\tilde T^B_{A;B} - \frac{1}{5}\tilde T_{;A}\biggl\}.
\end{eqnarray}
The space-time decomposition $D = 4 + 1$ leads to,
\begin{eqnarray}
\tilde R_{\mu\nu} &=& R_{\mu\nu} -  \frac{\phi_{;\mu;\nu}}{\phi}, \\
\tilde R_{ab} &=& \phi\Box\phi \\
\tilde R &=& R - 2\frac{\Box\phi}{\phi}.
\end{eqnarray}

The energy-momentum tensor $\tilde T_{AB}$ has the following components:
\begin{eqnarray}
\tilde T_{\mu\nu} &=& T_{\mu\nu}\\
\tilde T_{ab} &=& 0,\\
\tilde T &=& T = g^{\mu\nu} T_{\mu\nu}.
\end{eqnarray}

The field equations in four dimensions for $A = \mu$ and $B = \nu$ are as follows:
\begin{eqnarray}
R_{\mu\nu} - \frac{1}{5}g_{\mu\nu}R &=& 8\pi G\biggr\{T_{\mu\nu} - \frac{1}{5}g_{\mu\nu}T\biggl\} 
+ \frac{1}{\phi}\biggl\{\phi_{;\mu;\nu} - \frac{2}{5}g_{\mu\nu}\Box\phi\biggl\}.
\end{eqnarray}
Considering now the component $A = a, B = b$ of the multidimensional field equations, we obtain,
\begin{eqnarray}
\frac{\Box\phi}{\phi} = \frac{1}{3}(R - 8\pi G T).
\end{eqnarray}

The final set of equations can be written as follows:
\begin{eqnarray}
\label{e1}
R_{\mu\nu} &=& 8\pi G T_{\mu\nu} + \frac{1}{\phi}\biggr(\phi_{;\mu;\nu} - g_{\mu\nu}\Box\phi\biggl),\\
\label{e2}
\frac{\Box \phi}{\phi} &=& \frac{1}{3}\biggr(8\pi G T - R\biggl),\\
\label{e3}
8\pi G\biggr({T^\mu_\nu}_{;\mu} + \frac{\phi_{;\mu}}{\phi}T^\mu_\nu\biggl) &=& \frac{R_{;\nu}}{2}.
\end{eqnarray}
Eq. (\ref{e2}) is the trace of (\ref{e1}), while Eq. (\ref{e3}) comes from the Bianchi identity applied to (\ref{e1}). 

Let us now carry on the analysis of the effective equations (\ref{e1},\ref{e2},\ref{e3}). These equations have a very peculiar structure, 
which is similar to the Brans-Dicke gravity \cite{Brans} with $\omega = 0$, but with crucial difference that 
the Ricci scalar is absent in the right-hand side of the equation (\ref{e1}). The vacuum solutions display a singularity-free scenario which reveals instability due 
to the change of sign of the scalar field $\phi$ during the evolution of the universe \cite{FabrisKerner2025}.
In what follows, configurations with matter will be constructed with some very particular properties. In particular, 
the perturbative analysis with display important differences with respect to the vacuum case.

\section{Cosmology}

From now on the flat FLRW four-dimensional metric, describing an homogeneous and isotropic space, 
\begin{eqnarray}
ds^2 = dt^2 - a^2(dx^2 + dy^2 + dz^2).
\end{eqnarray}
is introduced in the field equations.
The matter content is described by a perfect fluid:
\begin{eqnarray}
T_{\mu\nu} = (\rho + p)u_\mu u_\nu - p g_{\mu\nu}.
\end{eqnarray}

Inserting the metric and the energy-momentum tensor in Eq. (\ref{e1}), two non-linear differential equations are obtained:
\begin{eqnarray}
3(\dot H + H^2) &=& - 8\pi G\rho - 3H\frac{\dot\phi}{\phi}, \\
\dot H + 3H^2 &=& 8\pi Gp + \frac{\ddot\phi}{\phi} + 2H\frac{\dot\phi}{\phi}.
\end{eqnarray}

There are three functions to be determined ($a$, $\rho$ and $\phi$). An additional hypothesis must be introduced. This hypothesis could mount to impose 
the conservation of the energy-momentum tensor, leading to the appearance of the cosmological constant as stressed previously. 
Here, however, another strategy will be explored, which leads to non-singular solutions, by imposing a constant Ricci scalar:
\begin{eqnarray}
\label{s-c}
R = - 6 (\dot H + 2H^2) = R_0 = \mbox{cte}.
\end{eqnarray}
This condition is satisfied if 
\begin{eqnarray}
a(t) = a_0 \cosh^{1/2} t,
\end{eqnarray}
with $R_0 = -3$. This solution represents a non-singular, eternal universe, exhibiting a bounce. Other solutions, with an initial singularity, 
are possible for positive and null values of $R_0$ but we will restrict ourselves to the non-singular configuration represented by the above solution with negative $R_0$. 
Notice that this solution remains valid regardless of the presence or absence of matter.

\section{Solutions with matter}

The complete set of equations equations in presence of matter is as follows:
\begin{eqnarray}
- 3(\dot H + H^2) &=& 8\pi G \rho - 3H \frac{\dot\phi}{\phi},\\
\dot H + 3 H^2 &=& 8\pi G p + \frac{\ddot \phi}{\phi} + 2H\frac{\dot\phi}{\phi}, \\
\dot\rho + 3H(\rho + p) + \frac{\dot\phi}{\phi}\rho &=& 0.
\end{eqnarray}

Supposing a linear, barotropic equation of state $p = \omega\rho$, with $\omega$ constant, the last equation has the solution,
\begin{eqnarray}
\rho = \frac{\rho_0}{\phi a^{3(1 + \omega)}}.
\end{eqnarray}

With this solution, the two remaining equations simplify to a ordinary, non-homogenous equation for $\phi$:
\begin{eqnarray}
\dot\phi - \biggr(\frac{\dot H}{H} + H\biggl)\phi = \frac{8\pi G}{3} \frac{\rho_0}{H a^{3(1 + \omega)}}.
\end{eqnarray}
This is a first order linear non-homogeneous equation for $\phi$, which can be solved once $a(t)$ and $\omega$ are given. 
This very peculiar property of the model results from the structure of equations (\ref{e1}-\ref{e3}) and the ansatz (\ref{s-c}). 

The solution of the homogenous part is given by,
\begin{eqnarray}
\phi_h =  A\frac{\sinh t}{\cosh^{1/2}t},
\end{eqnarray}
where $A$ is a constant.
To obtain the solution of the inhomogeneous equation, it is better to write,
\begin{eqnarray}
\phi = \dot a f,
\end{eqnarray}
leading to
\begin{eqnarray}
f = B \int \frac{dt}{\sinh^2 t \cosh^\frac{3\omega}{2} t},
\end{eqnarray}
with 
\begin{eqnarray}
B = \frac{16\pi G\rho_0}{3 a_0^{3(1 + \omega)}}.
\end{eqnarray}

The final solution is given by,
\begin{eqnarray}
\label{phi}
\phi =  A\frac{\sinh t}{\cosh^{1/2}t} + B \frac{\sinh t}{\cosh^{1/2}t}\int \frac{dt}{\sinh^2 t \cosh^\frac{3\omega}{2} t}
\end{eqnarray}

Now, some particular cases will be considered.

\subsection{Pressureless matter: $\omega = 0$}

In this case, the integral can be easily solved:
\begin{eqnarray}
\int \frac{dt}{\sinh^2 t} = - \coth t.
\end{eqnarray}
Hence, the solution for $\phi$ is given by,
\begin{eqnarray}
\label{s-d}
\phi = A\frac{\sinh t}{\cosh^{1/2}t} - B \cosh^{1/2}t.
\end{eqnarray}

The scalar field $\phi$ change of sign at the time $t_0$ satisfying the condition,
\begin{eqnarray}
\tanh t_0 = \frac{B}{A}.
\end{eqnarray}
Since $- 1 \leq \tanh t_0 \leq 1$, there is no change of sign in $\phi$ if $A \leq B$. In this case, $\phi$ is always negative, implying a repulsive gravity 
during all cosmic evolution. However, even though, the solution may be stable as it will be seen later.

\subsection{Domain wall: $\omega = - 2/3$}

Another case, is a fluid composed of domain wall topological defects \cite{vilenkin}. The equation of state is given by $p = - \frac{2}{3}\rho$. 
The integral in (\ref{phi}) can be easily done, leading to
\begin{eqnarray}
\label{dw}
\phi = A\frac{\sinh t}{\cosh^{1/2}t} - \frac{B}{\cosh^{1/2}t}.
\end{eqnarray}
There is a transition from positive to negative values if
\begin{eqnarray}
\sinh t_0 = \frac{B}{A}.
\end{eqnarray}
There is always a transition from positive to negative values for any value of $B$ and $A$.
If $A = 0$, $t_0 \rightarrow \infty$. However, as it happens for the pressureless fluid, $\phi$ is always negative.

\section{Stability}

In \cite{FabrisKerner2025} it has been shown that the singularity-free cosmological vacuum solutions coming from Eqs. (\ref{e1}-\ref{e3}) are unstable. 
Such instability appears in the evolution of gravitational waves, and the reason is connected with the change of sign of the scalar field $\phi$: 
during the evolution of the universe, $\phi$ is negative in the contracting phase, becoming positive in the expanding phase. 
The change of sign in $\phi$ coincides with the bouncing time, where there is a transition from contraction to expansion phases. 
Now, in presence of matter, there is the possibility to have no change of sign of $\phi$. However, if this case, $\phi$ may be purely negative 
at least for a pressureless and domain wall fluids investigate before. A complete perturbative analysis of the solutions found 
is out of the scope of the present text. However, we can show that there is no sign of instability for the tensorial modes and indications 
that instabilities are absent in scalar modes. We will discuss these cases in what follows.

\subsection{Tensorial modes}

The tensorial modes are obtained by introducing perturbations in Eqs (\ref{e1}-\ref{e3}) and by retaining only the traceless and divergence free 
of the metric perturbation $h_{ij}$. Here the synchronous gauge is used. 
For a throughout discussion of the gauge issue in UG, see Ref. \cite{suda}.

Using the expressions for the metric perturbations and of the energy-momentum tensor \cite{wei}, (see also \cite{FabrisKerner2025}) the evolution of the gravitational waves 
in presence of matter is given by,
\begin{eqnarray}
\label{gw-p}
\ddot h_{ij}  - \biggr(H - \frac{\dot\phi}{\phi}\biggl)\dot h_{ij} + \biggr\{\frac{k^2}{a^2} - 2 \biggr(\dot H + H^2 + H\frac{\dot\phi}{\phi}\biggl)\biggl\}h_{ij} = 0.
\end{eqnarray}
In this equation the wavenumber $k$ comes from the Fourier decomposition of the perturbations.

The evolution of the tensorial modes, given by Eq. (\ref{gw-p}) will now be analyzed for the two analytical solutions for the background determined previously.

\subsubsection{Pressureless fluid}

\begin{figure}[h]
    \centering
    \includegraphics[width=0.5\linewidth]{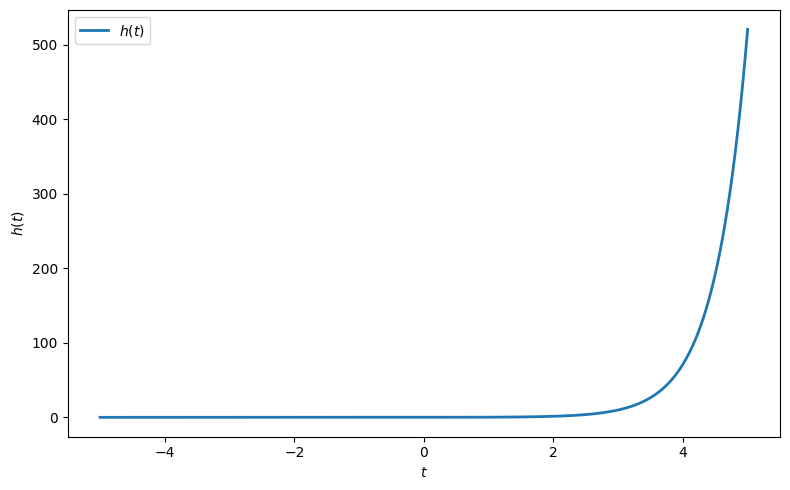}
    \caption{Evolution of the tensorial mode for Eq. (\ref{gw-p-p}), with $A = 0$ and $B = 1$, with the initial conditions at 
$t = - 5$, $h(t_0) = h'(t_0) = 10^{-6}$. The parameters $k$ and $a_0$ were fixed equal to 1.}
    \label{fig1}
\end{figure}

\begin{figure}[h]
    \centering
    \includegraphics[width=0.5\linewidth]{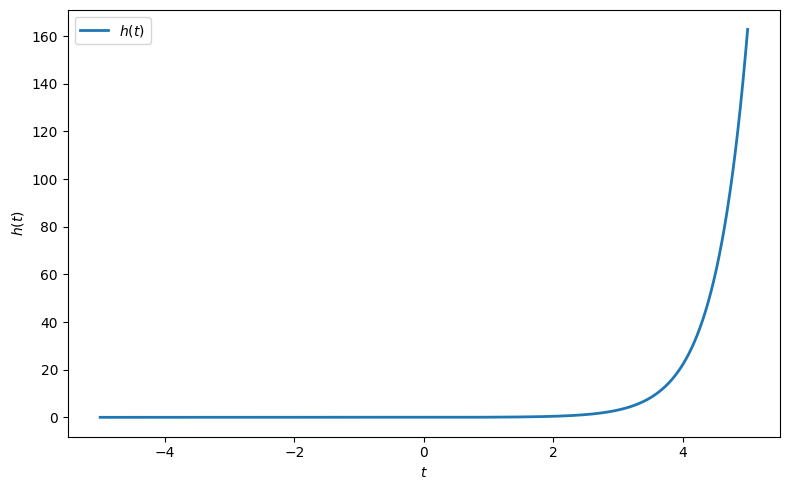}
    \caption{Evolution of the tensorial mode for Eq. (\ref{gw-p-p}), with $A = 0.5$ and $B = 1$, with the initial conditions at 
$t = - 5$, $h(t_0) = h'(t_0) = 10^{-6}$. The parameters $k$ and $a_0$ were fixed equal to 1.}
    \label{fig2}
\end{figure}

\begin{figure}[h]
    \centering
    \includegraphics[width=0.5\linewidth]{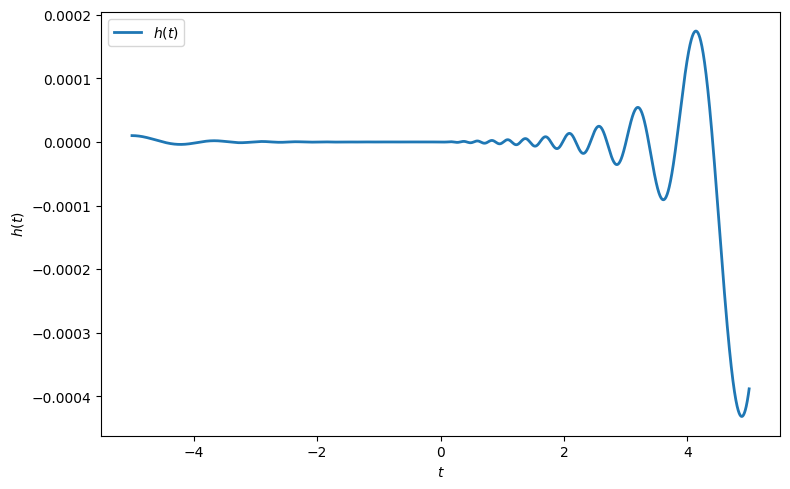}
    \caption{Evolution of the tensorial mode for Eq. (\ref{gw-p-p}), with $A = 0$ and $B = 1$, with the initial conditions at 
$t = - 5$, $h(t_0) = h'(t_0) = 10^{-6}$. It has been choosen $k = 30$ and $a_0 = 1$. Oscillations appears near the bounce.}
    \label{fig3}
\end{figure}

The simplest case is when $A = 0$ in (\ref{s-d}). This implies that,
\begin{eqnarray}
\label{c-p=0}
\phi = - B a \quad \rightarrow \quad \frac{\dot \phi}{\phi} = H.
\end{eqnarray}
Under such condition and using the relation (\ref{s-c}) with $R_0 = - 3$, equation (\ref{gw-p}) reduces to:
\begin{eqnarray}
\label{gw-p-p}
\ddot h_{ij} + \biggr\{\frac{k^2}{a^2} - 1\biggl\}h_{ij} = 0.
\end{eqnarray}
For $t \rightarrow \pm \infty$, the solutions take the form,
\begin{eqnarray}
h_{ij} = \epsilon_{ij}(c_1 e^{t} + c_2e^{-t}),
\end{eqnarray}
where $\epsilon_{ij}$ is the polarization tensor.
In order to have initially finite perturbations, as $t \rightarrow - \infty$, $c_2 = 0$. The perturbations grow but in the interval 
\begin{eqnarray}
- \cosh^{-1} \frac{k^2}{a^2_0} < t < + \cosh^{-1} \frac{k^2}{a^2_0},
\end{eqnarray}
they stop growing and begin to oscillate. The lower the scale ($k \rightarrow \infty$) the lower the amplification of the tensorial modes. 
In the other regime, for large scales, there is a an important amplification of the tensorial modes.
In the long wavelength limit $k \rightarrow 0$, the tensorial modes are amplified during all the evolution of the universe.
The general behavior for some specific configurations are given in figures \ref{fig1} and \ref{fig2}. The appearance of oscillations is exemplified in figure \ref{fig3}.

Even if the amplification may be huge for large scales perturbations, there is no instability in the sense discussed in Ref. \cite{FabrisKerner2025}: 
the perturbations remain always finite, diverging only asymptotically when the scale factor also diverges.

\subsubsection{Domain wall fluid}

The second case to be analyzed is the domain wall fluid. The solution for the $\phi$ is given by (\ref{dw}). If the arbitrary constant $A$ is again set equal to zero, 
the scalar field behaves as
\begin{eqnarray}
\phi = - \frac{B}{\cosh^{1/2}t} \propto - \frac{1}{a} \quad \rightarrow \quad \frac{\dot\phi}{\phi} = - H.
\end{eqnarray}
Hence the equation for the propagation of the tensorial modes reads,
\begin{eqnarray}
\label{gw-dw}
\ddot h_{ij}  - 2H\dot h_{ij} + \biggr\{\frac{k^2}{a^2} - 2 \dot H\biggl\}h_{ij} = 0.
\end{eqnarray}
Inserting the background solution, the equation becomes,
\begin{eqnarray}
\label{gw-dw-1}
\ddot h_{ij}  - \tanh t\dot h_{ij} + \biggr\{\frac{\tilde k^2}{\cosh t} -  \frac{1}{\cosh^2 t}\biggl\}h_{ij} = 0.
\end{eqnarray}

\begin{figure}[h]
    \centering
    \includegraphics[width=0.5\linewidth]{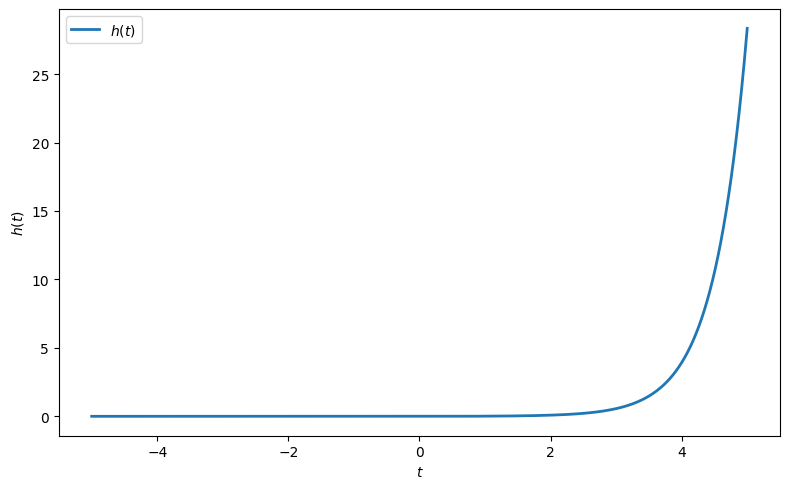}
    \caption{Evolution of the tensorial mode for Eq. (\ref{gw-dw-1}), with $A = 0$ and $B = 1$, with the initial conditions at $t = - 5$, $h(t_0) = h'(t_0) = 10^{-6}$. The parameters $k$ and $a_0$ were fixed equal to 1.}
    \label{fig4}
\end{figure}

The initial conditions can be obtained by inspecting the asymptotical solutions as $t \rightarrow - \infty$. The only finite solution is $h_{ij} \propto$ constant. 
The second term in Eq. (\ref{gw-dw-1}) implies a damping of the perturbations in the contraction phase and an anti-damping behavior during the expanding phase, 
leading to an amplification phenomenon. On the other hand, the last term implies that oscillations are presented in the beginning of the contraction phase 
and at the end of the expanding phase, and there are amplification near the bounce in the time interval,
\begin{eqnarray}
- \cosh^{-1} \frac{a_0^2}{k^2} < t < + \cosh^{-1} \frac{a_0^2}{k^2},
\end{eqnarray}
The amplification is less stronger than for the pressureless case. A specific example is given in figure \ref{fig4}
In any case, there is no possible divergence in a finite time interval.

\subsection{Scalar perturbations}

Even if the tensorial modes have a regular behavior, instabilities due to the violation of the energy conditions can be revealed by considering scalar perturbations. 
In General Relativity, for example, fluids models with negative squared sound velocity, even with regular behavior in the tensorial modes sector, may display strong instabilities 
in scalar modes, mainly those directly connected with density perturbations \cite{jerome}. However, for the model under analysis here the non-singular behavior of the background 
is not driven by an exotic fluid but by the peculiar geometric structure of the effective equations in four dimensions resulting from the dimensional reduction 
from the five-dimensional UG. 

A complete analysis of the scalar modes, and consequently of density perturbations, lies outside the purpose of the present article. One aspect that deserves a separate study 
concerns the ansatz of a constant four-dimensional Ricci scalar. This ansatz allow to obtain a simple expression for the divergence of energy-momentum tensor and, at the same time, 
an expression for the scale factor independent of the fluid. Even if the resulting equations for the background are all consistent with this ansatz, the perturbative analysis brings 
new questions. For example, does this ansatz must be also valid for a perturbed universe? Moreover, there are also some more fundamental questions concerning the status 
of the unimodular five-dimensional constraint in five dimension when projected to four dimensions, and the resulting impact on the gauge choice for the perturbed quantities, 
a problem already presented in the usual four-dimensional UG, see \cite{suda,bran}.

Now, we just sketch a simple, particular analysis of the scalar mode that suggests that the background configuration discussed above is stable. 

We postpone the complete perturbative analysis to a future work, mainly because there are many considerations that open the possibilities 
to more complex configurations as discussed above. Hence, we will consider the pressureless case, with the background given by (\ref{s-d}) with $A = 0$, 
which is not essential for the final results but simplify some expressions.

First, we note 
that the five-dimensional unimodular constraint, after reduction to four dimensions,
can be written as,
\begin{eqnarray}
\sqrt{-g}\phi = \mbox{constant}.
\end{eqnarray}
Perturbing this expression, using the synchronous gauge condition $h_{\mu0} = 0$, it results the relation,
\begin{eqnarray}
\frac{\delta\phi}{\phi} = \frac{h}{2}, \quad h = \frac{h_{kk}}{a^2}.
\end{eqnarray}
Perturbing the $0-0$ component of the (\ref{e1}), the final equation is,
\begin{eqnarray}
\ddot h + 4H\dot h - \frac{\nabla^2 h}{a^2} = 16\pi G\rho \delta, \quad \delta = \frac{\delta\rho}{\rho}.
\end{eqnarray}
The ansatz $R$ = cte is preserved at the perturbed level. This is one of the working hypothesis that may admit other variants. 
However, it is a reasonable and consistent hypothesis. One consequence is that, for pressureless matter, the rotational mode decouples 
and a simple expression can be obtained for the density perturbations.

 The perturbation of the resulting conservation law
\begin{eqnarray}
{T^\mu_\nu}_{;\mu} + \frac{\phi_{;\mu}}{\phi}T^\mu_\nu = 0,
\end{eqnarray}
after using the background equations, leads to,
\begin{eqnarray}
\dot\delta = 0, \quad \delta = \mbox{constant}.
\end{eqnarray}
Performing a Fourier decomposition, the final equation, for the pressureless case with the configuration (\ref{c-p=0}), is
\begin{eqnarray}
\ddot h + 4H\dot h + \frac{k^2 }{a^2}h = 16\pi G\rho \delta, \quad \delta = \frac{\delta\rho}{\rho}.
\end{eqnarray}
This equation, clearly, does not show any divergence, with a typical regular behavior of a damped, non-homogenous, oscillator.

Even if the computation has been only sketched, and a deeper analysis is required, including on the underlined hypothesis, the final result suggests 
the possibility of a stable behavior also for the scalar modes.

\section{Final Remarks}

The unimodular Kaluza-Klein gravity (KK-UG) after dimensional reduction to four dimensions leads to an effective theory similar to the Brans-Dicke one 
with $\tilde \omega = 0$, the important difference being though that the Ricci scalar is absent in the pure four-dimensional field equations. 
The use of the Bianchi identities implies the possibility that the energy-momentum tensor is not conserved separately, a usual property of Unimodular Gravity. 
In the vacuum case, the universe, under the ansatz imposing a constant Ricci scalar, the universe displays a bounce, avoiding singularity. 
It has been shown \cite{FabrisKerner2025} that such configuration is unstable. 

Similar instabilities appear already in the tensorial sector of the perturbed universe and are due to the change of sign of the scalar field 
(which, in the original theory, is connected with the fifth dimension). The inclusion of matter in the efective $4$-dimensional theory  
allows to obtain different scalar field configurations  without changing the behavior of the scale factor, remaining the same as in the vacuum case. 

This curious fact seems to be related to the particular geometric structure of the effective theory in four dimensions.
One feature of such solutions of matter is the possibility that the scalar field does not change its sign remaining, in certain cases, always negative. 
This changes drastically the behavior at perturbative level. In particular, the tensorial modes do not exhibit instabilities anymore, 
although the amplification of perturbations can be observed. An inspection of scalar perturbations also does not reveal instabilities, at least for a specific configuration.

Clearly, a complete perturbative analysis should be carried out. This implies the analysis of different variants of the perturbed model. 
For example, the condition $R = $ cte may be relaxed at the perturbative level, leading to a complex coupling of the perturbed quantities, at least in the scalar sector. 
The tensorial modes are not affected by relaxing or not the above constraint. A complete perturbative analysis is under study, including possible observational signatures 
related to the cosmological scenario described in the present work.

\bigskip

\noindent

{\bf Acknowledgements:} J.C.F. and S.F. thank CNPq (Brasil) and FAPES (Brasil) for partial financial support.

\end{document}